\documentclass[conference]{IEEEtran}
\IEEEoverridecommandlockouts

\usepackage{cite}
\usepackage{amsmath,amssymb,amsfonts}
\usepackage{algorithmic}
\usepackage{graphicx}
\usepackage{booktabs}
\usepackage{tabularx}  
\usepackage{textcomp}
\usepackage{xcolor}
\usepackage{subcaption}
\usepackage[font=small,labelfont=bf]{caption}
\usepackage[utf8]{inputenc}

\def\BibTeX{{\rm B\kern-.05em{\sc i\kern-.025em b}\kern-.08em
    T\kern-.1667em\lower.7ex\hbox{E}\kern-.125emX}}

\begin{document}

\title{Multistatic Radar Performance in the Presence of Distributed Wireless Synchronization  \\
\thanks{
Efforts sponsored by the U.S. Government under the Training and Readiness Accelerator II (TReX II), OTA. The U.S. Government is authorized to reproduce and distribute reprints for Governmental purposes, notwithstanding any copyright notation thereon. The views and conclusions contained herein are those of the authors and should not be interpreted as necessarily representing the official policies or endorsements, either expressed or implied, of the U.S. Government.}
\vspace{-0.65em}
}

\author{ 
\IEEEauthorblockN{
         Kumar Sai Bondada\IEEEauthorrefmark{1},
         Daniel J. Jakubisin\IEEEauthorrefmark{1}\IEEEauthorrefmark{2},
         R. Michael Buehrer\IEEEauthorrefmark{1}   
} 
     \IEEEauthorblockA{
         \IEEEauthorrefmark{1}Wireless@VT, Bradley Department of ECE, Virginia Tech, Blacksburg, VA, USA \\
         \IEEEauthorrefmark{2}Virginia Tech National Security Institute, Blacksburg, VA, USA  
         } 
                  \vspace{-3.0 em}
         }
\maketitle

\begin{abstract}
This paper proposes a multistatic radar (MSR) system utilizing a distributed wireless synchronization protocol. The wireless synchronization protocol uses a two-tone waveform exchange for frequency synchronization and a bi-directional waveform exchange for time synchronization, independent of GPS. A Bayesian Cramér–Rao lower bound (BCRLB) framework is developed to quantify the impact of synchronization offsets on joint delay and Doppler estimation, and consequently, on target localization and velocity estimation accuracy. Simulation results derived from the analytical expressions establish the extent to which the residual synchronization offsets degrade the MSR's performance.
The performance of the synchronization links primarily depends on the synchronization-link channel and  transmit parameters; optimizing these parameters enables the MSR configuration to surpass the monostatic performance and approach the ideal case. Furthermore, the simulated synchronization-link parameters suggest that practical implementation is feasible.
\end{abstract}

\begin{IEEEkeywords}
Multistatic radar, distributed sensing, Bayesian CRLB, wireless frequency and time synchronization
\vspace{-0.75em}
\end{IEEEkeywords}

\section{Introduction}
Multistatic radars~\cite{9455149,OHAGAN2018253} (MSRs) offer significant advantages in detecting stealth targets that are designed to scatter energy away from the monostatic radar line-of-sight (LOS). Passive mode operation makes them inherently more covert and resilient to electronic countermeasures. The growing deployment of unmanned aerial vehicles (UAVs) and vehicular nodes further enhances the appeal of MSR for modern sensing and surveillance applications. However, their performance critically depends on achieving precise synchronization and maintaining accurate knowledge of the cooperating radar locations. Prior research has explored various synchronization strategies, including wired~\cite{9289008}, GPS (GNSS)-based~\cite{7944470,8835507,Beasley2023GNSSDO}, and GPS-independent wireless approaches~\cite{RFclock,9994246,9443078}. 

In GPS-based approaches, an oscillator drives the digital subsystems, such as sampling-clock PLLs (ADC/DACs), carrier-generating PLLs (LO), and data clocks, and is disciplined by a GPS timing receiver, commonly referred to as a GPS-disciplined oscillator (GPSDO). These units provide stable timing and frequency references, with typical timing accuracies on the order of nanoseconds and fractional frequency stability on the order of $10^{-13}$–$10^{-11}$~\cite{Beasley2023GNSSDO, article_Lombardi}, achieving long-term stability. The short-term stability is determined by the oscillator used and its phase-noise characteristics. However, GPS signals can be easily spoofed or jammed, require LOS visibility to satellites, and typically need a long acquisition time to achieve lock. Moreover, high-performance GPSDOs are expensive, although low-cost commercial off-the-shelf (COTS) options are available with moderate stability.

To overcome the dependency on GPS, independent wireless synchronization approaches have been developed, which can be broadly classified into centralized \cite{RFclock,9443078,9994246} and decentralized \cite{10734842} architectures. These are designed for general distributed wireless communication systems, with a focus on improving synchronization performance and demonstrated sub-hertz frequency accuracy and nanosecond-to-picosecond (ns–ps) time accuracy. In centralized systems, a leader–follower architecture is adopted, where the followers synchronize to the leader by exchanging frequency and time reference signals. Frequency synchronization techniques typically rely on the wireless transmission of an analog \cite{7218555,RFclock,9994246,bondada2025experimentaldemonstrationrobustdistributed} or digitally modulated reference from the leader. Time synchronization, on the other hand, is achieved through bi-directional waveform (BDW) exhanges \cite{RFclock,9443078,9994246, 6289099, 6624252, 6854706}, which can be implemented using either timestamp-based or timestamp-free protocols. A key challenge for frequency synchronization is its susceptibility to multi-path and Doppler effects, while time synchronization schemes require symmetric propagation delays between the leader and followers. Furthermore, the achievable accuracy is strongly dependent on the SNR of the wireless link.


Prior studies have investigated GPS-based synchronization for MSR from different perspectives. Coherent network radar operation using custom GPSDOs was demonstrated in~\cite{7944470}. The follow-up work~\cite{8835507} analyzed synchronization limitations and proposed LOS phase compensation for bistatic coherence. The work in~\cite{Beasley2023GNSSDO} evaluated commercial GPSDOs for practical radar synchronization. Wireless adaptation of the white rabbit precision time protocol (WR-PTP) was studied in \cite{Rico2015,Gilligan2020}, with a particular focus on synchronization. In contrast, this paper investigates the performance of a leader–follower MSR architecture employing GPS-independent wireless synchronization. A Bayesian Cramér–Rao lower bound (BCRLB)~\cite{1165144, 382422, 8321430} framework is developed to study the impact of the residual synchronization offsets on target position and velocity estimation accuracy. In this work, frequency synchronization is achieved by extracting a reference clock from a two-tone waveform (TTW) broadcasted by leader to all followers~\cite{7218555, RFclock, 9994246, bondada2025experimentaldemonstrationrobustdistributed}. Time synchronization is accomplished using a timestamp-free bidirectional waveform exchange that estimates and corrects clock offsets while compensating for propagation delays~\cite{6289099, 6624252, 6854706}. Simulation results demonstrate the MSR performance under multiple configurations of synchronization-link channel and transmit parameters, and compared with the monostatic radar.


The paper is structured as follows. First, the system model incorporating synchronization offsets is introduced. Next, the BCRLB framework is presented, where the synchronization offsets are treated as nuisance parameters with prior distributions. The subsequent section describes the MSR geometry and derives the corresponding CRLBs for target position and velocity. The procedures for frequency and time synchronization are then outlined. Finally, the simulation results are discussed, emphasizing the synchronization integrated MSR's performance, followed by the conclusion and future works.

\begin{figure}[h!]
    \centering
    \includegraphics[width=\linewidth]{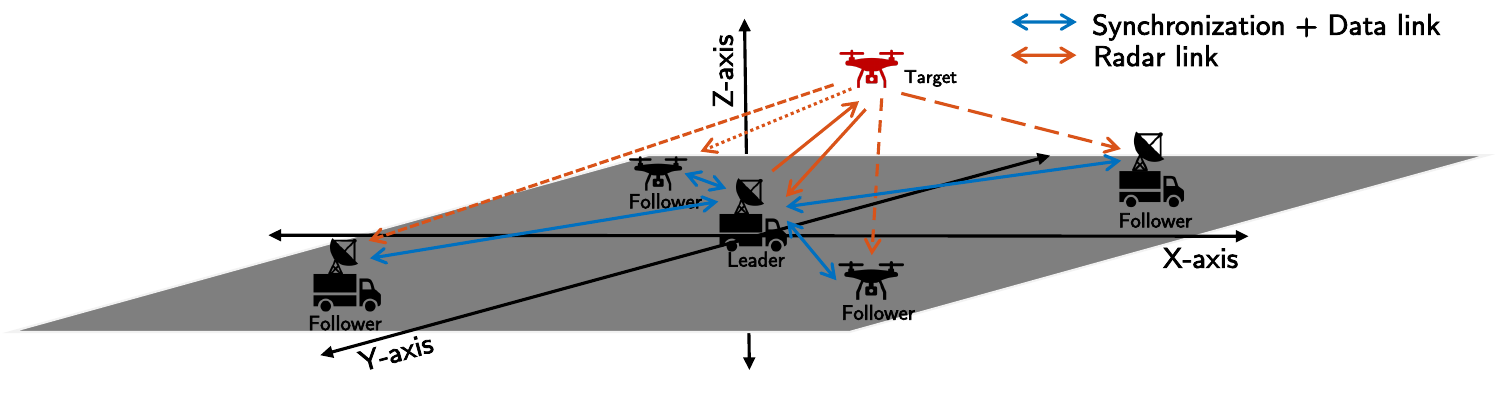}
    \caption{Multistatic radar system illustrating a leader radar and multiple follower radars.}
    \label{fig:multi_static_radar_arch}
    \vspace{-1em}
\end{figure}




\textbf{Notation:} Boldface uppercase letters denote matrices; boldface lowercase letters denote vectors.

\vspace{-1mm}

\section{Signal Model with Synchronization Offsets}
In this section, we present the signal model incorporating synchronization offsets along with the underlying assumptions. Fig.~\ref{fig:multi_static_radar_arch} illustrates a MSR architecture featuring inter-radar wireless links for synchronization and data exchange, as well as radar links between the radars and the target. The followers communicate wirelessly with the leader to achieve synchronization and to exchange Time of Arrival (TOA) and Frequency of Arrival (FOA) measurements. The radars performing measurements are assumed to be either mobile or fixed wireless nodes with known locations relative to the leader. The focus of this work is on the impact of synchronization errors, while the position uncertainty of the followers is neglected. The followers, considered as passive radars (e.g., mounted on drones or vehicular nodes), are geographically distributed around the leader, providing scalability in the number of nodes and adaptability to mobility based on operational requirements. 


The transmitted signal from the leader radar is expressed as
$
s_p(t) = s(t)e^{j 2\pi f_c t},
$
where $s(t)$ denotes the complex baseband radar pulse with a duration of $T$ seconds, and $f_c$ is the carrier frequency. 
Assuming a LOS channel between the transmitter and target as well as receiver and target and neglecting synchronization offsets, the received signal at the $i$th follower after down-conversion can be written as

\begin{equation}
r_i(t) = \sqrt{G_i}\, s(t - \tau_i)e^{j\left(2\pi f_i (t - \tau_i) + \theta_i\right)},
\end{equation}
where $G_i$ accounts for the large-scale path loss and the radar cross-section (RCS) of the target, $\tau_i$ is the propagation delay, $f_i$ is the Doppler shift observed at the $i$th radar due to target motion, and $\theta_i$ is the phase shift caused by the reflection.


In practical systems, synchronization is typically performed parallel or in advance to target detection, to make sure that the \textit{carrier frequency offsets} (CFOs) and \textit{clock (time) offsets} (TOs) are corrected, ensuring carrier and timing alignment among all cooperating radars. However, residual CFOs and TOs may still remain, resulting in frequency and timing mismatches across the radars. Incorporating these residual offsets into the ideal received signal and accounting for complex additive white Gaussian noise (AWGN), $n_i(t) \sim \mathcal{CN}(0, N_0)$, the received signal at the $i$th radar is
\begin{equation}
r_i(t) = \sqrt{G_i}\, s(t - \tau_i - \delta t_i) e^{j \left( 2\pi ( f_i + \delta f_i ) (t - \tau_i) + \theta_i \right)} + n_i(t),
\end{equation}
where $\delta t_i$ and $\delta f_i$ represent the residual TOs and CFOs, respectively and will be treated as nuisance parameters~\cite{891222}. We combine the amplitude term $\sqrt{G_i}$ and the constant phasor $h_i = h_i^{r} + j h_i^{i} = e^{j(-2\pi (f_i + \delta f_i) \tau_i + \theta_i)}$ into a single complex channel coefficient $A_i \triangleq \sqrt{G_i} h_i$. The received signal can thus be expressed as
\small
\begin{equation}
r_i(t) = A_i\, s\!\left(t - (\tau_i + \delta t_i)\right) e^{j 2\pi (f_i + \delta f_i) t} + n_i(t).
\end{equation}
\normalsize
The terms $G_i$ and $h_i$ are included in the effective SNR, and since $|h_i|^2 = 1$, the received SNR is given by $\gamma_i =  G_i \frac{ E_s}{N_0}$. The deterministic parameters of interest are the TOA, $\tau_i$, and the FOA, $f_i$. The residual TOs and CFOs are modeled as independent Gaussian random variables, $\delta t_i \sim \mathcal{N}(0, \sigma_{t_i}^2)$ and $\delta f_i \sim \mathcal{N}(0, \sigma_{f_i}^2)$, with variances determined by the performance of the employed wireless synchronization techniques. The deterministic mean of the received (noise-free) signal is then given by $\mu_i(t) \triangleq \mathbb{E}\{r_i(t)\} = A_i\, s\!\left(t - (\tau_i + \delta t_i)\right) e^{j 2\pi (f_i + \delta f_i) t}$.
\vspace{-0.25em}



\section{BCRLB for Delay $\tau$ and Doppler $f_d$}
We present the lower bounds on the variances of delay and Doppler estimation in the presence of synchronization offsets. The BCLRB~\cite{1165144, 382422, 8321430} provides a lower bound on the mean squared error (MSE) of a estimator when prior information about the unknown parameters is available. It extends the classical CRLB by incorporating information from both the likelihood and the parameters prior. Let $\boldsymbol{\theta} \in  \mathbb{R}^p$ denote a random parameter vector of interest, and let $\boldsymbol{x} \in  \mathbb{R}^N$ denote the observation vector obtained from a statistical model parameterized by $\boldsymbol{\theta}$, with $\hat{\boldsymbol{\theta}}(\boldsymbol{x})$ the estimator of $\boldsymbol{\theta}$. The BCRLB is defined as
\begin{equation}
\mathbb{E}_{\boldsymbol{\theta},\boldsymbol{x}}
\!\left[
(\hat{\boldsymbol{\theta}}(\boldsymbol{x}) - \boldsymbol{\theta})
(\hat{\boldsymbol{\theta}}(\boldsymbol{x}) - \boldsymbol{\theta})^\top
\right]
\succeq
\mathbf{I}_B^{-1},
\end{equation}
where $\mathbf{I}_B \in \mathbb{R}^{p \times p}$ is the Bayesian information matrix (BIM), defined as
\begin{equation}
\mathbf{I}_B
\triangleq
\mathbb{E}_{\boldsymbol{\theta},\boldsymbol{x}}
\!\left[
\nabla_{\boldsymbol{\theta}} \log p(\boldsymbol{x},\boldsymbol{\theta})\,
\nabla_{\boldsymbol{\theta}}^\top \log p(\boldsymbol{x},\boldsymbol{\theta})
\right].
\end{equation}
The $I_B$ can be decomposed into two components,
$ \mathbf{I}_P + \mathbf{I}_N,
$
where $\mathbf{I}_P$ represents the prior information term, given by
\begin{equation}
\mathbf{I}_P
\triangleq
\mathbb{E}_{\boldsymbol{\theta}}
\!\left[
\nabla_{\boldsymbol{\theta}} \log p(\boldsymbol{\theta})\,
\nabla_{\boldsymbol{\theta}}^\top \log p(\boldsymbol{\theta})
\right],
\end{equation}
and $\mathbf{I}_N$ denotes the Fisher information matrix (FIM) averaged over the parameters prior,
\begin{equation}
\mathbf{I}_N
\triangleq
\mathbb{E}_{\boldsymbol{\theta},\boldsymbol{x}}
\!\left[
\nabla_{\boldsymbol{\theta}} \log p(\boldsymbol{x}|\boldsymbol{\theta})\,
\nabla_{\boldsymbol{\theta}}^\top \log p(\boldsymbol{x}|\boldsymbol{\theta})
\right]
= \mathbb{E}_{\boldsymbol{\theta}}\!\left[\mathbf{I}_F(\boldsymbol{\theta})\right],
\end{equation}
with FIM as
$
\mathbf{I}_F(\boldsymbol{\theta})
\triangleq
\mathbb{E}_{\boldsymbol{x}|\boldsymbol{\theta}}
\!\left[
\nabla_{\boldsymbol{\theta}} \log p(\boldsymbol{x}|\boldsymbol{\theta})\,
\nabla_{\boldsymbol{\theta}}^\top \log p(\boldsymbol{x}|\boldsymbol{\theta})
\right].
$

\vspace{-1em}
\subsection{BCRLB for Delay $\tau_i$ and Doppler $f_i$}

\subsubsection{Fisher Information}
The FIM for the parameter vector $\boldsymbol{\theta} = [\tau_i, f_i, \delta t_i, \delta f_i]$ is defined as
\[
\mathbf{I}_F^{i}(\boldsymbol{\theta}) =
\begin{bmatrix}
I_{\tau_i \tau_i}      & I_{\tau_i f_i}      & I_{\tau_i \delta t_i}      & I_{\tau_i \delta f_i} \\
I_{f_i \tau_i}         & I_{f_i f_i}         & I_{f_i \delta t_i}         & I_{f_i \delta f_i}    \\
I_{\delta t_i \tau_i}  & I_{\delta t_i f_i}  & I_{\delta t_i \delta t_i}  & I_{\delta t_i \delta f_i} \\
I_{\delta f_i \tau_i}  & I_{\delta f_i f_i}  & I_{\delta f_i \delta t_i}  & I_{\delta f_i \delta f_i}
\end{bmatrix}.
\]
For parameters $a_i$ and $b_i$, and the mean signal $\mu_i(t)$ observed in complex AWGN with a two-sided power spectral density (PSD) of $N_0$, the FIM elements \cite{10.5555/151045} are given by
\small
\begin{equation}
\label{eq:FI_base}
I_{a_i b_i}
= \frac{2}{N_0} \, \Re \!\int_{-\infty}^{\infty}
\frac{\partial \mu_i(t)}{\partial a_i}
\frac{\partial \mu_i^*(t)}{\partial b_i}\, dt.
\end{equation}
\normalsize
\subsubsection{Partial Derivatives of the Mean}
The partial derivatives of the mean signal $\mu_i(t)$ with respect to the parameters are obtained as
\small
\begin{align*}
\frac{\partial \mu_i(t)}{\partial \tau_i}
&= -A_i\, \frac{\partial s(t - (\tau_i + \delta t_i))}{\partial \tau_i}\, e^{j 2\pi (f_i + \delta f_i) t}, \\
\frac{\partial \mu_i(t)}{\partial \delta t_i}
&= -A_i\, \frac{\partial s(t - (\tau_i + \delta t_i))}{\partial \delta t_i}\, e^{j 2\pi (f_i + \delta f_i) t}, \\
\frac{\partial \mu_i(t)}{\partial f_i}
&= j 2\pi t\, A_i\, s(t - (\tau_i + \delta t_i))\, e^{j 2\pi (f_i + \delta f_i) t}, \\
\frac{\partial \mu_i(t)}{\partial \delta f_i}
&= j 2\pi t\, A_i\, s(t - (\tau_i + \delta t_i))\, e^{j 2\pi (f_i + \delta f_i) t}.
\end{align*}
\normalsize
Since the mean $\mu_i(t)$ depends on the composite parameters $(\tau_i + \delta t_i)$ and $(f_i + \delta f_i)$, these parameters are inseparable. Therefore, the FIM becomes singular due to collinear columns.






\subsubsection{Full FIM with Nuisance Parameters}
Since $\frac{\partial \mu_i}{\partial \tau_i} = \frac{\partial \mu_i}{\partial \delta t_i}$ and $\frac{\partial \mu_i}{\partial f_i} = \frac{\partial \mu_i}{\partial \delta f_i}$,  
the FIM can be written as
$
\mathbf{I}_F^{i} =
\begin{bmatrix}
\mathbf{A}^i & \mathbf{A}^i \\
\mathbf{A}^i & \mathbf{A}^i
\end{bmatrix},
$
where $\mathbf{A}^i$ denotes the FIM corresponding to the delay–Doppler parameters.
 In this work, the transmitted radar waveform is assumed to be a sequence of linear frequency modulated (LFM) pulses. The delay–Doppler FIM expressions derived in~\cite{923295} are used directly to avoid re-derivation. The FIM is given by
\small
\begin{equation}
\mathbf{A}^i
= 2 P \mathrm{SNR}_i
\begin{bmatrix}
 \frac{1}{3} \pi^2 f_B^2   & 
-\frac{1}{6} \pi f_B T_0 \\
-\frac{1}{6} \pi f_B T_0 & \frac{1}{12} T_0^2 
\left( 1 + \left( \frac{T_R}{T_0} \right)^2 (P^2 - 1) \right)
\end{bmatrix}
\end{equation}
\normalsize
where
$
\mathrm{SNR}_i = \gamma_i \, \mathbf{a}(\phi)^{*} 
\mathbf{N}_{\Sigma}^{-1} \mathbf{a}(\phi)$ is the (post-processed \cite{9994246}) SNR for a single pulse. Here, $P$ denotes the number of pulses, $T_0$ is the single-pulse duration, $f_B$ is the bandwidth of the chirp pulse, $T_R$ is the pulse repetition interval, $\mathbf{a}(\phi)$ is the array response vector, and $\mathbf{N}_{\Sigma}$ represents the spatial noise covariance matrix. Here, $\mathbf{I}_N^{i} = \mathbf{I}_F^{i}$, as $\mathbf{I}_F^{i}$ does not depend on the parameters of interest.

\subsubsection{Adding Gaussian Priors for Offsets}
The prior information matrix adds information only for the offset parameters \((\delta t_i, \delta f_i)\) (delay and Doppler are unknown deterministic) and is expressed as
\small
\begin{equation}
\mathbf{I}_P^{i}
= \operatorname{diag}(0, 0, \sigma_{t_i}^{-2}, \sigma_{f_i}^{-2})
= 
\begin{bmatrix}
0 & 0 \\
0 & \boldsymbol{\Lambda}^i
\end{bmatrix},
\,
\boldsymbol{\Lambda}^i =
\begin{bmatrix}
\sigma_{t_i}^{-2} & 0 \\
0 & \sigma_{f_i}^{-2}
\end{bmatrix}.
\end{equation}
\normalsize
Accordingly, the BIM is given by
\small
\begin{equation}
\mathbf{I}_B^{i}
= \mathbf{I}_D^{i} + \mathbf{I}_P^{i}
= 
\begin{bmatrix}
\mathbf{A}^i & \mathbf{A}^i \\
\mathbf{A}^i & \mathbf{A}^i + \boldsymbol{\Lambda}^i
\end{bmatrix}.
\end{equation}
\normalsize
\subsubsection{Equivalent BIM for $\begin{bmatrix}
    \tau_i & f_i
\end{bmatrix}^\top$} Eliminating the nuisance parameters using the Schur complement yields
$
\mathbf{I}^{\text{eq}}_{i}
= \mathbf{A}^i - \mathbf{A}^i(\mathbf{A}^i + \boldsymbol{\Lambda}^i)^{-1}\mathbf{A}^i.
$

\subsubsection{Bayesian CRLB}
The lower bound for estimation error for the parameters $
    \tau_i$ and $f_i$ is obtained as 
\small
\begin{equation}
\operatorname{CRLB}(\begin{bmatrix}
    \tau_i & f_i
\end{bmatrix}^\top ) 
= \left(\mathbf{I}^{\text{eq}}_{i}\right)^{-1}.
\label{eq:crlb_tau_f}
\end{equation}
\normalsize




\section{CRLB for Position $\mathbf{p}$ and Velocity $\mathbf{v}$}

We consider the leader as the monostatic transmitter–receiver radar (indexed as $i=0$) and $N$ receive-only followers ($i = 1, \dots, N$). The target position and velocity are denoted by $\mathbf{p} \in \mathbb{R}^n$ and $\mathbf{v} \in \mathbb{R}^n$, respectively, where $n$ is 3. The transmitter is located at $\mathbf{p}_t$, receiver $i$ is located at $\mathbf{p}_{r,i}$, and the monostatic radar satisfies $\mathbf{p}_t = \mathbf{p}_{r,0}$. The target-to-radar unit vectors and corresponding ranges are defined as
\small
\begin{align}
\hat{\mathbf{u}}_{t} &\triangleq \frac{\mathbf{p} - \mathbf{p}_t}{\|\mathbf{p} - \mathbf{p}_t\|}, 
& r_t &\triangleq \|\mathbf{p} - \mathbf{p}_t\|, \\
\hat{\mathbf{u}}_{r,i} &\triangleq \frac{\mathbf{p} - \mathbf{p}_{r,i}}{\|\mathbf{p} - \mathbf{p}_{r,i}\|}, 
& r_{r,i} &\triangleq \|\mathbf{p} - \mathbf{p}_{r,i}\|.
\end{align}
\normalsize
Each $i$th radar produces a joint TOA and FOA observation given by
\small
\begin{equation}
\mathbf{x}_i \triangleq 
\begin{bmatrix}
T_i \\
F_i
\end{bmatrix}
=
\underbrace{
\begin{bmatrix}
\tau_i(\mathbf{p}) \\
f_i(\mathbf{p}, \mathbf{v})
\end{bmatrix}
}_{\boldsymbol{\mu}_i(\boldsymbol{\psi})}
+
\underbrace{
\begin{bmatrix}
w_i \\
v_i
\end{bmatrix}
}_{\mathbf{n}_i},
\label{eq:y_i}
\end{equation}
\normalsize
where $\boldsymbol{\psi} = [\mathbf{p}^\top \ \mathbf{v}^\top]^\top$ represents the parameters of interest. The terms $w_i$ and $v_i$ denote the TOA and FOA measurement errors, respectively. The noise covariance for $\mathbf{n}_i \sim \mathcal{N}(0, \boldsymbol{\Sigma}_i)$ is determined from~\eqref{eq:crlb_tau_f}, i.e.,
$
\boldsymbol{\Sigma}_i = \operatorname{CRLB}([\tau_i, f_i]),
$
which accounts for the both measurements and residual frequency and time offsets.

The deterministic mean functions for the TOA and FOA measurements are given by
\small
\begin{align}
\tau_i(\mathbf{p}) &=
\begin{cases}
\dfrac{r_t + r_{r,i}}{c}, & i \ge 1 \quad \text{(bistatic)} \\[4pt]
\dfrac{2r_t}{c}, & i = 0 \quad \text{(monostatic)}
\end{cases}\!, 
\label{eq:tau} \\
f_i(\mathbf{p}, \mathbf{v}) &=
\begin{cases}
\dfrac{1}{\lambda}\!\left(\hat{\mathbf{u}}_{t} + \hat{\mathbf{u}}_{r,i}\right)^{\!\top}\!\mathbf{v}, & i \ge 1 \\
\dfrac{2}{\lambda}\,\hat{\mathbf{u}}_{r,0}^{\top}\!\mathbf{v}, & i = 0
\end{cases}\!,
\label{eq:foa}
\end{align}
\normalsize
where $\lambda$ denotes the radar wavelength. For the bistatic case ($i \ge 1$), the TOA corresponds to the total propagation delay from the leader to the target and then to follower $i$, whereas for the monostatic case ($i = 0$), it represents the round-trip delay. Similarly, the FOA expressions correspond to the Doppler shifts induced by the relative radial velocity $\mathbf{v}$ components of the target with respect to each radar.


Stacking the measurement and mean vectors 
$
\mathbf{x} = [\mathbf{x}_0^\top, \dots, \mathbf{x}_{N}^\top]^\top,$ 
and $
\boldsymbol{\mu} = [\boldsymbol{\mu}_0^\top, \dots, \boldsymbol{\mu}_{N}^\top]^\top $,
and since the measurements at each radar are statistically independent, the overall noise covariance matrix becomes a block-diagonal matrix $\mathbf{R} = \mathrm{blkdiag}(\boldsymbol{\Sigma}_0, \dots, \boldsymbol{\Sigma}_{N})$.
The full FIM can then be written as the sum of the contributions from each radar
\small
\begin{align}
\mathbf{I}(\boldsymbol{\psi})
&= 
\frac{\partial \boldsymbol{\mu}}{\partial \boldsymbol{\psi}}
\mathbf{R}^{-1}
\frac{\partial \boldsymbol{\mu}}{\partial \boldsymbol{\psi}} ^{\!\top}\notag 
= 
\sum_{i=0}^{N}
\frac{\partial \boldsymbol{\mu}_i}{\partial \boldsymbol{\psi}}
\boldsymbol{\Sigma}_i^{-1}
\frac{\partial \boldsymbol{\mu}_i}{\partial \boldsymbol{\psi}}^{\!\top}
.
\end{align}
\normalsize
\subsection{Partial Derivatives of the Mean}
For the $i$th radar, the derivative of the mean vector with respect to the parameter vector $\boldsymbol{\psi}$ is given by $\frac{\partial \boldsymbol{\mu}_i}{\partial \boldsymbol{\psi}} = \mathbf{G}_i$,
where
\small$
\mathbf{G}_i =
\begin{bmatrix}
\dfrac{\partial \tau_i}{\partial \mathbf{p}} & \dfrac{\partial f_i}{\partial \mathbf{p}}  \\
 \dfrac{\partial \tau_i}{\partial \mathbf{v}} & \dfrac{\partial f_i}{\partial \mathbf{v}}
\end{bmatrix}
\in \mathbb{R}^{2n \times 2}.$ \normalsize
 The derivatives $\frac{\partial \boldsymbol{\mu}_i}{\partial \boldsymbol{\psi}}$ are derived using the definitions in Appendix~\ref{subsec:appendix-derivs}.  
For a general unit vector defined as $\hat{\mathbf{u}}(\mathbf{p}) = \frac{\mathbf{p} - \mathbf{a}}{\|\mathbf{x} - \mathbf{a}\|}$ with range $r = \|\mathbf{p} - \mathbf{a}\|$, the following key derivative identities hold
\small
\begin{align}
\frac{\partial r}{\partial \mathbf{p}} &= \hat{\mathbf{u}}^\top, \qquad
\frac{\partial \hat{\mathbf{u}}}{\partial \mathbf{p}} = \frac{1}{r}\!\left(\mathbf{I} - \hat{\mathbf{u}}\hat{\mathbf{u}}^\top\right).
\label{eq:key-identities}
\end{align}
\normalsize
From~\eqref{eq:tau} and~\eqref{eq:key-identities}, the TOA derivatives with respect to position and velocity are given by
\small
\begin{align}
\frac{\partial \tau_i}{\partial \mathbf{p}} &=
\begin{cases}
\dfrac{1}{c}\!\left(\hat{\mathbf{u}}_{t} + \hat{\mathbf{u}}_{r,i}\right)^{\!\top}, & i \ge 1 \\
\dfrac{2}{c}\,\hat{\mathbf{u}}_{r,0}^{\top}, & i = 0
\end{cases}\!,
\qquad
\frac{\partial \tau_i}{\partial \mathbf{v}} = \mathbf{0}^{\top}. \notag
\end{align}\normalsize
From~\eqref{eq:foa}, the FOA derivatives with respect to velocity are
\small
\begin{align}
\frac{\partial f_i}{\partial \mathbf{v}} &=
\begin{cases}
\dfrac{1}{\lambda}\!\left(\hat{\mathbf{u}}_{t} + \hat{\mathbf{u}}_{r,i}\right)^{\!\top}, & i \ge 1 \\
\dfrac{2}{\lambda}\,\hat{\mathbf{u}}_{r,0}^{\top}, & i = 0
\end{cases}\!, \notag 
\end{align}
\normalsize
and using $\frac{\partial \hat{\mathbf{u}}}{\partial \mathbf{p}}$ from~\eqref{eq:key-identities}, the FOA derivatives with respect to position are expressed as
\small
\begin{align}
\frac{\partial f_i}{\partial \mathbf{p}} &=
\begin{cases}
\dfrac{1}{\lambda}\,
\mathbf{v}^{\!\top}\!\left[
\dfrac{1}{r_t}\!\left(\mathbf{I} - \hat{\mathbf{u}}_{t}\hat{\mathbf{u}}_{t}^\top\right)
+ \dfrac{1}{r_{r,i}}\!\left(\mathbf{I} - \hat{\mathbf{u}}_{r,i}\hat{\mathbf{u}}_{r,i}^\top\right)
\right], & i \ge 1 \\
\dfrac{2}{\lambda}\,
\mathbf{v}^{\!\top}\!\left[
\dfrac{1}{r_{r,0}}\!\left(\mathbf{I} - \hat{\mathbf{u}}_{r,0}\hat{\mathbf{u}}_{r,0}^\top\right)
\right], & i = 0 
\end{cases}\!. \notag
\end{align}
\normalsize
Finally, the FIM contribution from the $i$th radar reduces to
$
\mathbf{G}_i^{\top}\boldsymbol{\Sigma}_i^{-1}\mathbf{G}_i,
$
and the total effective FIM for the target position and velocity parameters is obtained by summing the individual contributions 
and the lower bound on the estimation error for parameters $\mathbf{p}$ and $\mathbf{v}$ is given by
\small
\begin{equation}
\operatorname{CRLB}(\begin{bmatrix}
    \mathbf{p} & \mathbf{v}
\end{bmatrix}^\top ) 
= \mathbf{J}(\boldsymbol{\psi})^{-1}.
\label{eq:CRLB_x_p}
\end{equation}
\normalsize

\section{Distributed Wireless Synchronization}
In this section, we briefly describe wireless synchronization employed by the leader and followers.

\subsubsection{Frequency Synchronization}
The leader broadcasts a TTW, where the tones are separated by a fixed reference frequency \cite{7218555, RFclock, 9994246, bondada2025experimentaldemonstrationrobustdistributed}. Each follower extracts the reference clock signal using a dedicated circuit and uses it as a reference for clock synchronization. The TTW generated at the leader is also based on the same reference frequency. This ensures that the clocks and carriers at all radars tick at the same rate. However, clock offsets may still remain unaligned. 

\subsubsection{Time Synchronization} Clock (absolute time) alignment is achieved through a BDW exchange, following the method described in \cite{6289099, 6624252, 6854706}. In this process, follower transmits a known reference waveform to the leader. The leader then retransmits the reference waveform at a designated transmit time such that its clock tick is centered between the reception and transmission instants. The follower receives this waveform, estimates its clock offset with respect to the leader, and applies the necessary correction. Readers are referred to the cited works for mathematical details and additional insights related to synchronization.


\section{Simulation Results}
\begin{figure}[ht]
    \centering
    \includegraphics[width=0.85\linewidth]{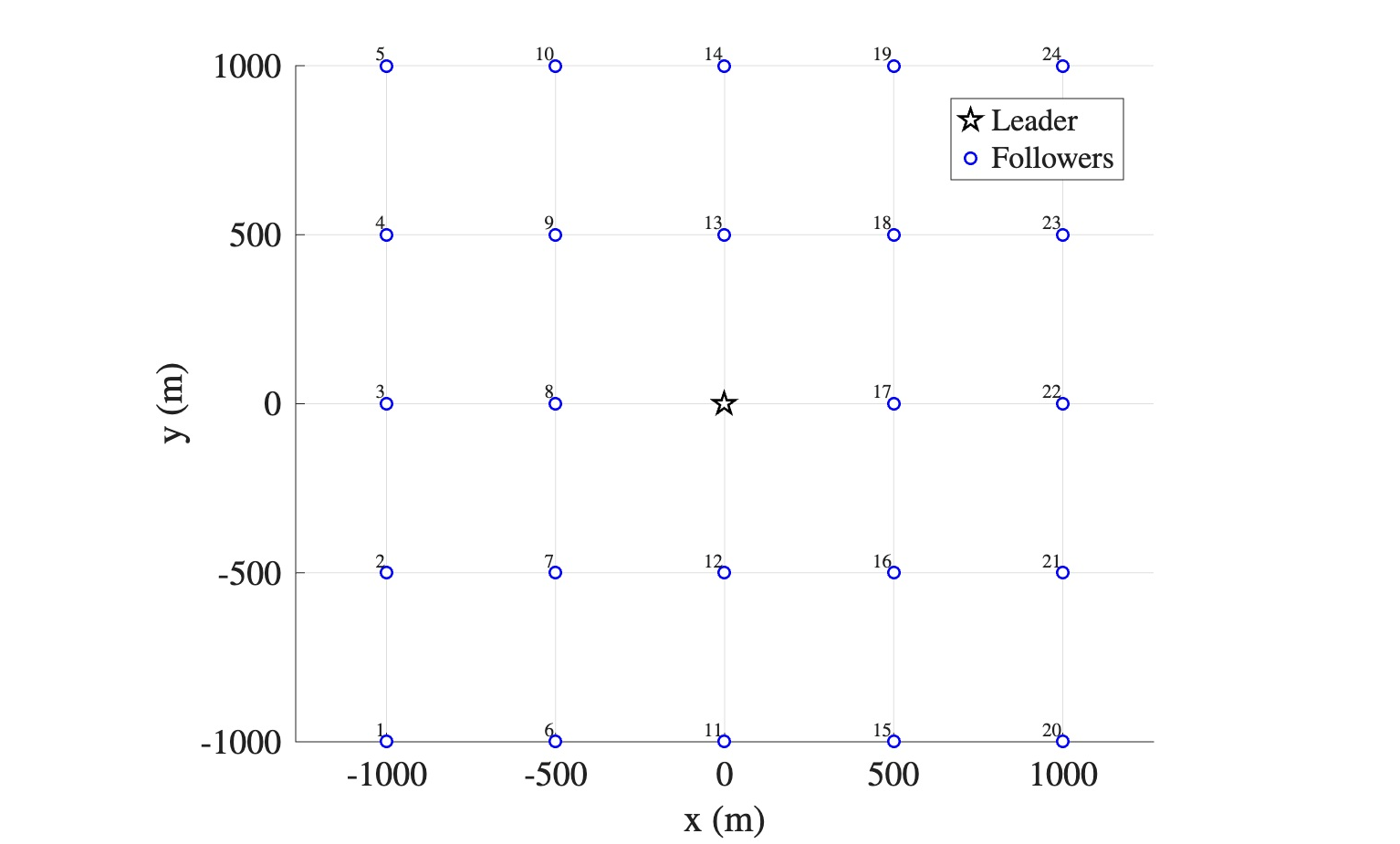}
    \caption{Two-dimensional layout of the leader and follower radars.}
    \label{fig:layout}
    \vspace{-1em}
\end{figure}
In this section, we evaluate the performance of the MSR system with and without follower radars. The simulation setup, illustrated in Fig.~\ref{fig:layout}, consists of equally spaced one leader and 24 followers (along with follower index) each equipped with single antenna (resulting in total 25 antennas). A free-space path-loss model is assumed with a radar cross section (RCS) of \(0.1~\text{m}^2\). The key radar parameters are summarized in Table~\ref{table:radar_params}.


\begin{table}[!t]
\caption{Radar Simulation Parameters}
\label{table:radar_params}
\centering
\begin{tabular}{l c}
\toprule
\textbf{Parameter} & \textbf{Value} \\
\midrule
Number of radars (leader + followers) & 25 \\
Inter-radar spacing, \(d\) & 500 m \\
 Leader transmit power, \(P_{\text{tx}}\) & 33 dBm \\
 Transmit / Receive antenna gains, \((G_t, G_r)\) & (10, 10) dB \\
 Signal (sweep) bandwidth, \(f_B\) & 50 MHz \\
 Pulse duration, \(T_p\) & 100~\textmu s \\
 Number of pulses, \(P\) & 100 \\
Receiver noise figure & 4 dB \\
Carrier frequency & 2.4 GHz \\
\bottomrule
\end{tabular}
\vspace{-1em}
\end{table}

\begin{table}[!t]
\caption{Synchronization Parameters}
\label{table:sync_params}
\centering
\begin{tabular}{l c}
\toprule
\textbf{Parameter} & \textbf{Value} \\
\midrule
Leader / Follower transmit power, \(P^L_{\text{sync}}/P^F_{\text{sync}}\) & 23/23 dBm \\
 Transmit / Receive antenna gains, \((G_t, G_r)\) & (1, 1) dB \\
 Carrier frequency & 2.4 GHz \\
\bottomrule
\end{tabular}
\vspace{-1em}
\end{table}

For the leader–follower synchronization links, an Urban Macro (UMa) large-scale path-loss model with Rician fading is adopted~\cite{7414180}, capturing both dominant LOS and diffuse NLOS components (exponential delay power profile). Depending on the node heights, the channels are categorized as Air-to-Ground (A2G) or Air-to-Air (A2A). As the radar altitude increases, the LOS component becomes stronger, leading to higher SNR and reduced multi-path effects. A Rician factor of 3 is used based on~\cite{7414180}, with both leader and followers placed at a height of 10~m (A2A configuration). For frequency synchronization, the leader broadcasts a TTW with 10~MHz tone separation. Each follower extracts a reference clock from the received waveform and uses it to drive its internal circuitry. The frequency difference between the extracted reference and the nominal 10~MHz tone spacing is estimated and multiplied by the PLL (Phase-locked loop) factor to obtain the CFO error. The PLL factor refers to the multiplication ratio that a PLL applies to its reference input frequency to generate the carrier. For time synchronization, followers initiate a bidirectional waveform exchange at a transmit power of \( P_{\text{sync}}^F \), and the leader responds accordingly. Since clock alignment depends on delay estimation (independent of the TOA estimation used for target tracking), the clock estimation error is modeled as twice the delay estimation error, accounting for estimation at both ends. An LFM waveform is used for delay estimation \cite{9173801}. The CRLB for the first-path delay estimate is

\small
\begin{equation}
    \mathrm{var}\!\left(\widehat{\tau}_{\mathrm{sync}}\right)
    \ge
    \frac{1}{2\,\zeta_f^{2}\,\mathrm{SNR}_{\mathrm{sync}}B_{\mathrm{sync}}T_{\mathrm{sync}}\,|h^1_{\mathrm{sync}}|^{2}},
\end{equation} 
\normalsize
where $\zeta_f^{2}$ is the mean-squared bandwidth, the
(pre-processed) $\mathrm{SNR}_{\mathrm{sync}}=P_{\mathrm{sync}}/\sigma_{n}^{2}$ with transmit
power $P_{\mathrm{sync}}$ and noise variance $\sigma_{n}^{2}$, and $|h^1_{\mathrm{sync}}|^{2}$ is
the dominant-path channel gain. For an LFM pulse of bandwidth $B_{\mathrm{sync}}$,
$
    \zeta_f^{2}= \frac{(\pi B_{\mathrm{sync}})^{2}}{3}.
$
Thus, the standard deviations of the frequency $\delta f$ and clock time $\delta t$ errors are computed and used as priors in the FIM calculation for delay and Doppler estimation at each radar.
A time-bandwidth product of $B_{\mathrm{sync}}T_{\mathrm{sync}} = 128$ is assumed. 

\begin{figure}[htpb]
    \centering
    \begin{subfigure}{\columnwidth}
        \centering
        \includegraphics[width=0.81\columnwidth]{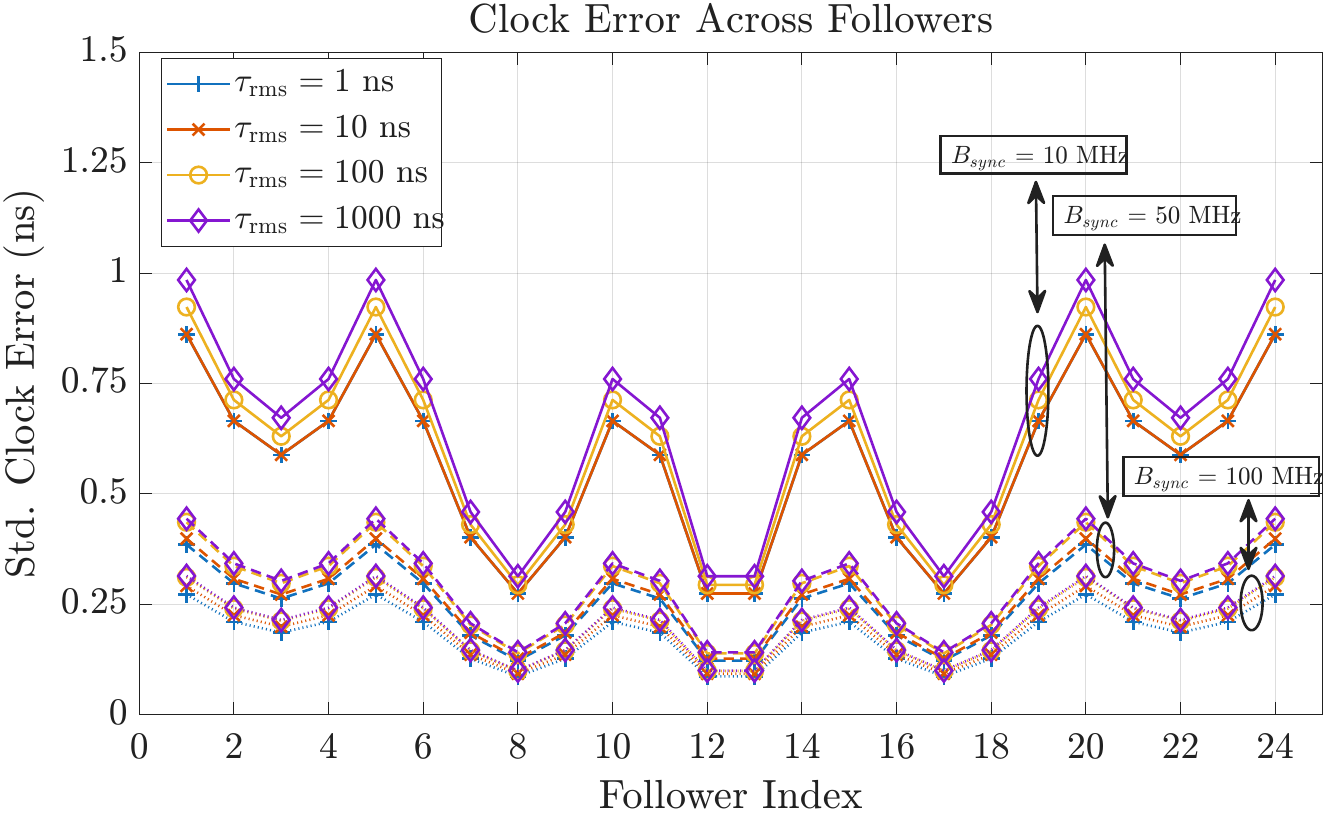}
        \caption{Std. Deviation of Clock Errors (in ns).}
        \label{fig:timeErrors}
    \end{subfigure}
    \hfill
    \begin{subfigure}{\columnwidth}
        \centering
        \includegraphics[width=0.81\columnwidth]{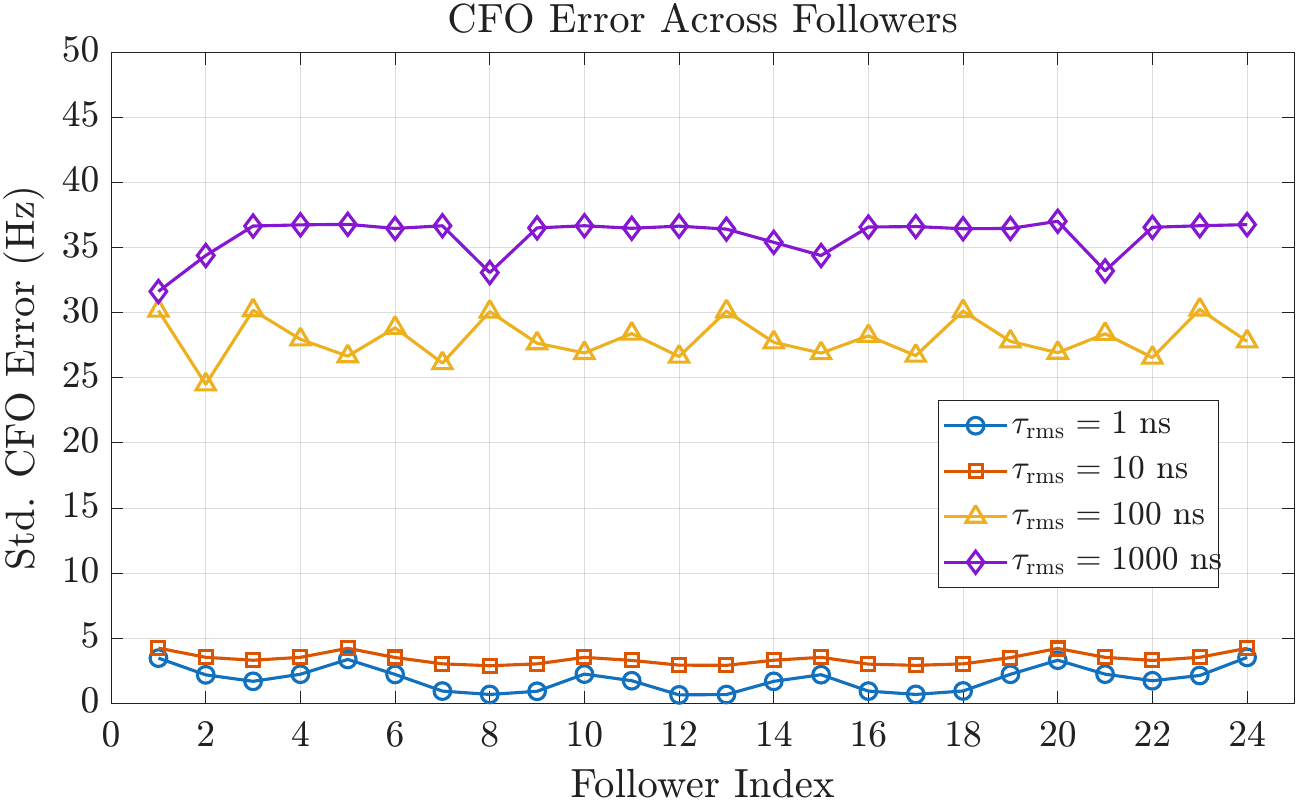}
        \caption{Std. Deviation of CFO Errors (in Hz)).}
        \label{fig:cfo}
    \end{subfigure}
    \caption{Time and frequency synchronization performance: (a) bidirectional waveform exchange for clock alignment and (b) two-tone waveform broadcasting for frequency alignment.}
    \label{fig:time_freq_sync_performance}
    \vspace{-1em}
\end{figure}

The standard deviations of the CFO errors and clock error are shown in the Fig.~\ref{fig:time_freq_sync_performance}. The maximum Doppler shift in the synchronization links is considered to be 0.1~Hz, arising from environmental disturbances. The performance plots show the clock error in nanoseconds (ns), with higher errors observed for followers farther from the leader and lower errors for those closer to it, indicating that higher SNR yields smaller clock errors. Increasing the $B_{\mathrm{sync}}$ further reduces the clock error. The CFO errors are primarily influenced by Doppler shifts and diffuse multi-path components in the environment. As multi-path severity increases, the likelihood of one or both tones experiencing deep fades also rises. Lower delay spreads correspond to higher coherence bandwidths. Hence, higher CFO errors are observed for delay spreads exceeding 100~ns.

\begin{table}[!t]
\centering
\caption{Target Simulation Parameters}
\label{tab:table_target_sync}
\begin{tabular}{l c}
\toprule
\textbf{Parameter} & \textbf{Value (Min, Max)} \\
\midrule
Target location (X-axis) [m] & (-1000, 1000) \\
Target location (Y-axis) [m] & (-1000, 1000) \\
Target height [m] & (50, 100) \\
Target velocity [m/s] & (0, 20) \\
\bottomrule
\end{tabular}
\vspace{-2em}
\end{table}

\begin{figure}[htpb]
    \centering
    \begin{subfigure}{\columnwidth}
        \centering
        \includegraphics[width=0.85\columnwidth]{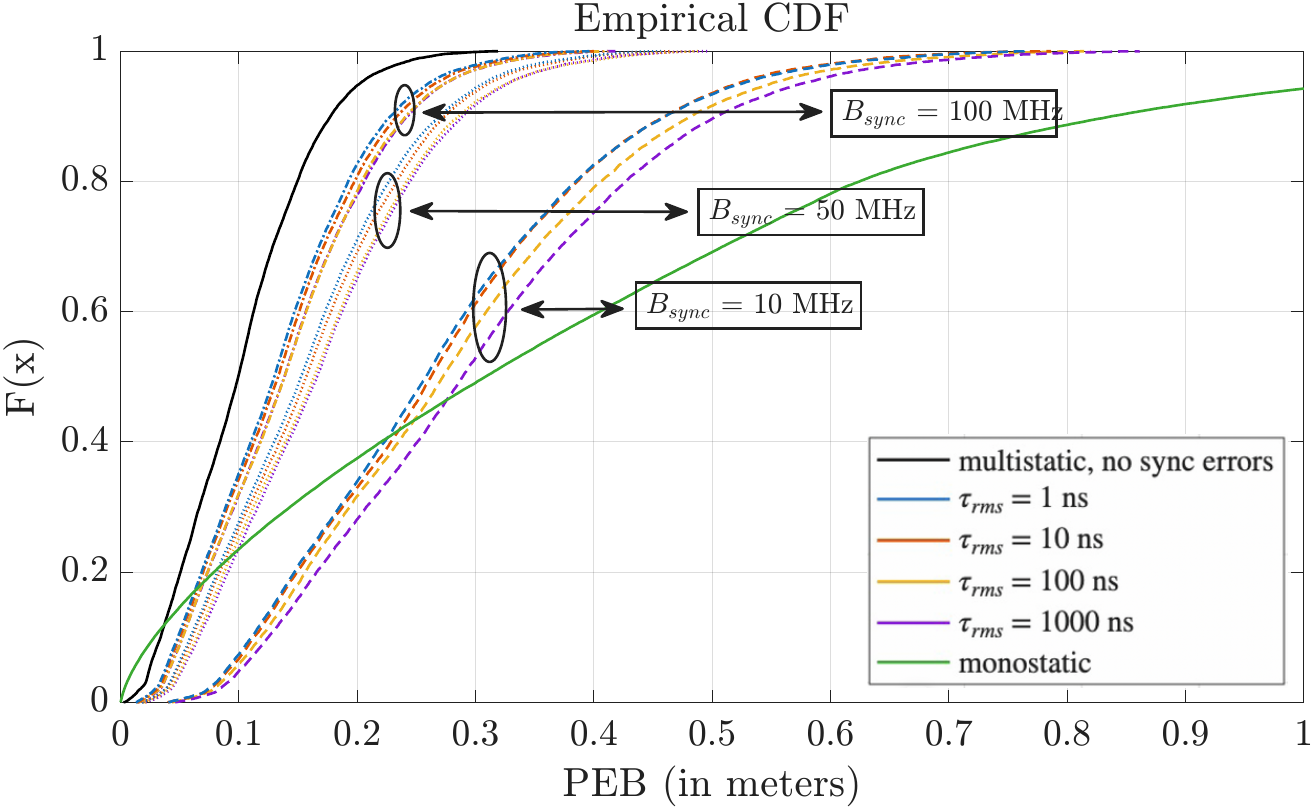}
        \caption{Position error bound (PEB).}
        \label{fig:peb}
    \end{subfigure}
    \hfill
    \begin{subfigure}{\columnwidth}
        \centering
        \includegraphics[width=0.85\columnwidth]
        {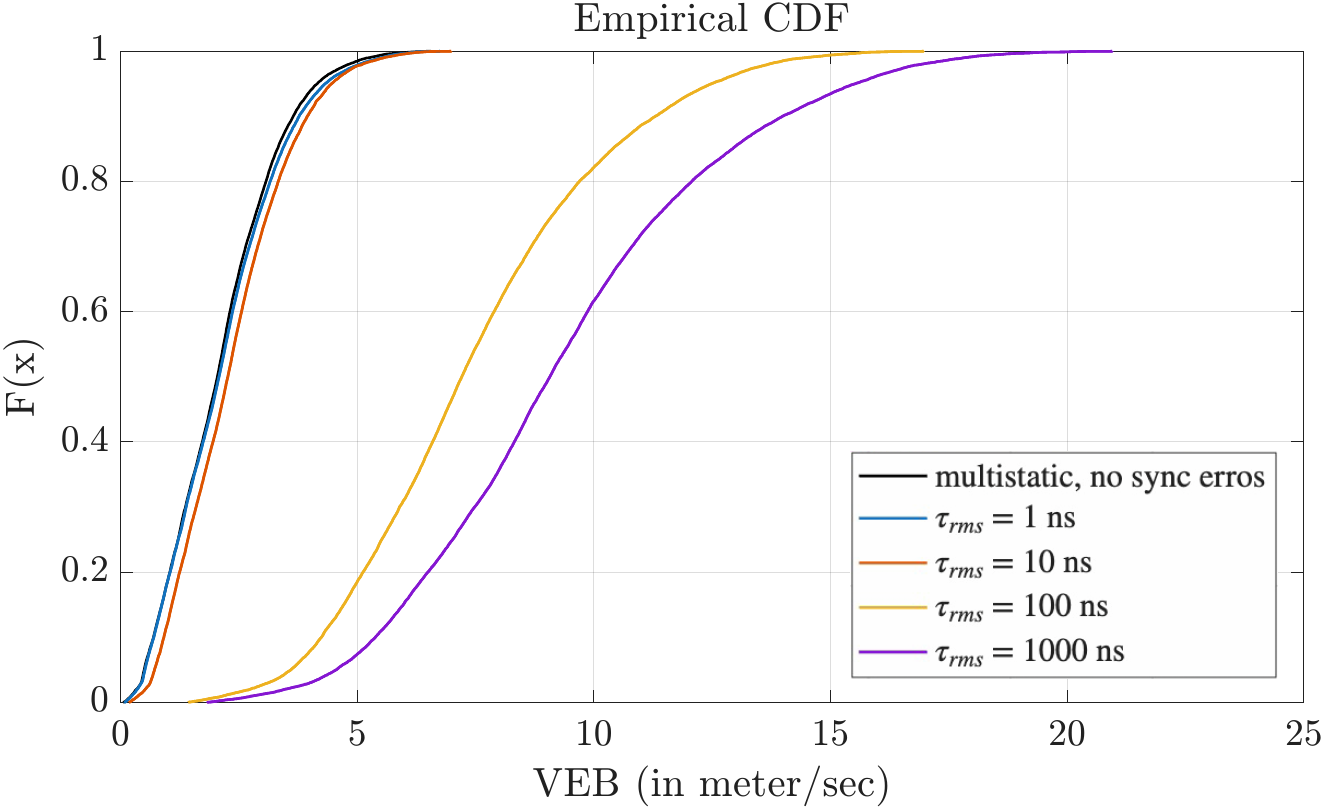}
        \caption{Velocity error bound (VEB).}
        \label{fig:veb}
    \end{subfigure}
    \caption{Multistatic radar performance under ideal conditions (no synchronization offsets) and with synchronization offsets.}
    \label{fig:perf_comparison}
    \vspace{-1em}
\end{figure}

Fig.~\ref{fig:perf_comparison} compares the MSR performance under ideal conditions and with synchronization offsets introduced by the integrated synchronization technique. The PEB (position error bound) and VEB (velocity error bound) are obtained from \eqref{eq:CRLB_x_p}, computed as the square root of the first three and the next three diagonal terms, respectively, with target locations and velocities uniformly drawn from the ranges in Table~\ref{tab:table_target_sync}. Performance improves with increasing $B_{\mathrm{sync}}$ used for delay estimation, while stronger diffuse multi-path (increasing delay spread $\tau_{rms}$) degrade accuracy. As synchronization-link performance improves, the MSR approaches the ideal case. The PEB is also compared with that of the monostatic radar~\cite{923295} (equipped with antenna array of \(3 \times 3 \times 3\) antenna array, resulting in total of 27 antenna) for fair comparison. The VEB remains unaffected by the synchronization bandwidth, as it is only influenced by the CFO errors resulting from frequency synchronization.

Clock errors propagate into TOA measurements, degrading target position estimation, while CFO errors translate into FOA measurements, reducing velocity accuracy. The plots confirm these effects, showing that increased synchronization errors directly deteriorate PEB and VEB. Improving delay estimation in the synchronization links enhances clock offset correction, allowing the MSR to approach ideal performance. Delay estimation accuracy increases with higher radar altitudes (stronger LOS) and greater transmit power, whereas frequency synchronization is primarily limited by maximum Doppler shifts and diffuse multi-path, both of which can be mitigated by increasing radar height. Under strong LOS conditions, a Doppler difference of 1~Hz between two followers results in a CFO error of approximately 240~Hz at a 2.4~GHz carrier (PLL factor of 240 for a 10 MHz reference input to PLL).

\section{Conclusion \& Future Works}
This paper presented a performance analysis framework for a wirelessly synchronized leader-follower MSR system, where frequency and time synchronization were achieved using two-tone waveform and bi-directional waveform exchanges, respectively. The BCRLB framework was derived to quantify the impact of synchronization offsets on target position and velocity estimation accuracy. Simulation results showed that synchronization imperfections lead to a noticeable degradation in performance, however, the MSR still outperforms the monostatic configuration. \textit{Favorable channel conditions enhance synchronization links, leading to optimal synchronization performance and, consequently, improved target tracking}. Beyond the presented results, this work establishes a foundation for low-altitude transceiver deployments, facilitating distributed radar and joint communication–sensing applications. Relevant extensions of this work include comparison with GPSDO solutions, and modeling of phase noise. 
Furthermore, this work suggests that joint optimization of follower radar placement is needed to ensure reliable target detection while maintaining synchronization link performance.


\section*{Appendix: Vector Derivatives}
\label{subsec:appendix-derivs}
Let $\mathbf{r}(\mathbf{p})=\mathbf{x}-\mathbf{a}$ and $r(\mathbf{p})=\|\mathbf{r}\|$.
\paragraph*{(i) Range derivative}
From $r^2=\mathbf{r}^\top\mathbf{r}$,
$
2r\,\mathrm{d}r = 2\,\mathbf{r}^\top\mathrm{d}\mathbf{p}
\Rightarrow \mathrm{d}r=\frac{\mathbf{r}^\top}{\|\mathbf{r}\|}\mathrm{d}\mathbf{p},
$
so $\frac{\partial r}{\partial \mathbf{p}}=\hat{\mathbf{u}}^\top$ with $\hat{\mathbf{u}}=\mathbf{r}/r$.

\paragraph*{(ii) Unit-vector Jacobian}
With $\hat{\mathbf{u}}=\mathbf{r}/r$,
$
\mathrm{d}\hat{\mathbf{u}} = \frac{1}{r}\,\mathrm{d}\mathbf{r} - \frac{\mathbf{r}}{r^2}\,\mathrm{d}r
= \left[\frac{1}{r}\mathbf{I}-\frac{1}{r}\hat{\mathbf{u}}\hat{\mathbf{u}}^\top\right]\mathrm{d}\mathbf{x},
$
hence $\frac{\partial \hat{\mathbf{u}}}{\partial \mathbf{p}}=\frac{1}{r}\!\left(\mathbf{I}-\hat{\mathbf{u}}\hat{\mathbf{u}}^\top\right)$.

\bibliographystyle{IEEEtran}
\bibliography{reference}

\end{document}